\documentclass[preprint,3p]{elsarticle}

\usepackage{amssymb}
\usepackage{url}
\usepackage{physics}

\newcommand{\me}[0]{\mathrm{e}}
\newcommand{\unit}[1]{\;\mathrm{#1}}

\usepackage[pdftex,
       colorlinks=true,
       urlcolor=blue,       % \href{...}{...} external (URL)
        bookmarksopen=true]{hyperref}

\newcommand{\npb}{{Nucl.\ Phys.\ B }}

\newcounter{bla}

\journal{Computer Physics Communications}

\renewcommand{\vec}[1]{\boldsymbol{#1}}
\def\eps{\varepsilon}

\begin{document}

\begin{frontmatter}

\title{{\tt AAfrag 2.01:} Interpolation routines for
  Monte Carlo results on
  secondary production including light antinuclei in hadronic interactions
  }

\author[a]{M.~Kachelrie\ss}
\author[b,c]{S.~Ostapchenko}
\author[a]{J.~Tjemsland\corref{author}}
\cortext[author] {Corresponding author}
\address[a]{Institutt for fysikk, NTNU, Trondheim, Norway}
\address[b]{II. Institute for Theoretical Physics, Hamburg University, Hamburg,
  Germany}
\address[c]{D.V. Skobeltsyn Institute of Nuclear Physics, Moscow State
  University, Moscow, Russia}

\begin{abstract}
Light antinuclei, like antideuteron and antihelium-3, are ideal probes for new,
exotic physics because their astrophysical backgrounds are suppressed at low
energies. In order to exploit fully the inherent discovery potential of light
antinuclei, a reliable description of their production cross sections in
cosmic ray interactions is crucial. We provide therefore the cross sections
of antideuteron and antihelium-3  production in $pp$, $p$He, He$p$, HeHe,
$\bar pp$ and $\bar p$He collisions at energies relevant for secondary
production in the Milky Way, in a tabulated form which is convinient to use.
These predictions are based on QGSJET-II-04m
and the state of the art coalescence model WiFunC, which evaluates the
coalesence probability on an event-by-event basis, including both momentum
correlations and the dependence on the emission volume. 
In addition, we comment on the importance of a Monte Carlo description of the 
antideuteron production and on the use of event generators in general.
In particular, we discuss the effect of two-particle
momentum correlations provided by Monte Carlo event generators on antinuclei
production.
\end{abstract}


\end{frontmatter}

\noindent
{\bf PROGRAM SUMMARY}

\begin{small}
\noindent
{\em Program Title:} {\tt AAfrag 2.01} \\
%{\em CPC Library link to program files:} (to be added by Technical Editor) \\
{\em Developer's repository link:} \url{https://aafrag.sourceforge.io} \\
%{\em Code Ocean capsule:} (to be added by Technical Editor)\\
{\em Licensing provisions:} CC BY-NC 4.0 \\
{\em Programming language:} The program exists both in a Python 3 and
  Fortran 90 version\\
{\em Supplementary material:}                                 \\
{\em Journal reference of previous version:}
  Comp. Phys. Comm. 245, 106846 (2019) \\
{\em Does the new version supersede the previous version?:} Yes   \\
{\em Reasons for the new version:}
  Inclusion of antinuclei tables, and a Python 3 version for the
  pedestrian \\
{\em Summary of revisions:} Improved formatting of tables, inclusion of new
  secondaries, inclusion of Python version \\
{\em Nature of problem:} Calculation of secondaries
  (photons, neutrinos, electrons, positrons, protons,
  antiprotons, antideuterons and antihelium-3)
  produced in hadronic interactions\\
{\em Solution method:}
  Results from the Monte Carlo simulation QGSJET-II-04m are interpolated\\
%\begin{thebibliography}{0}
%\bibitem{1}Reference 1
%\end{thebibliography}

\end{small}
  
\section{Introduction}

A precise knowledge of the production cross section of secondaries in hadronic
interactions is important in many applications in astrophysics and astroparticle
physics, ranging from deducing the cosmic ray spectrum from the observation
of secondary photons~\cite{Neronov:2011wi,Fermi-LAT:2012edv,Kachelriess:2012fz}
to indirect dark matter searches
using cosmic ray antinuclei~\cite{Donato:1999gy,vonDoetinchem:2020vbj}. 
Such studies are either based on convenient parametrisations tuned to available
accelerator data, or on Monte Carlo generators. The former have to rely on
empirical
scaling laws, which are unreliable when extrapolated outside the measured
kinematical range. Furthermore, such parametrisations are typically provided
only for protons, and therefore a ``nuclear enhancement factor'' 
has to be used to describe the production cross sections in interactions
involving nuclei.
However, nuclear enhancement factors are, especially close to threshold,
not able to capture the intricate dependence on the energy and the nuclear
mass number of these cross
sections~\cite{Kachelriess:2014oma,Kachelriess:2014mga,Kachelriess:2020uoh}.
Monte Carlo event generators, on the other hand, are generally 
less convenient for a user, but can describe consistently both hadron--hadron
and hadron--nucleus collisions. In Ref.~\cite{Kachelriess:2019ifk}, the
production cross sections for photons, neutrinos, electrons, positrons, (anti-)
protons, and (anti-) neutrons in various interactions relevant for secondary
production in the Milky Way were therefore derived from the event generator
QGSJET-II-04m~\cite{Ostapchenko:2010vb,Ostapchenko:2013pia,Kachelriess:2015wpa}
and made publicly available in an easily usable form.
In addition, Ref.~\cite{Koldobskiy:2021nld} discussed the
differences to other published parametrisations of the
photon, neutrino, electron, and positron spectra and provided a python
version of the interpolation subroutines.

The differences between parametrisations and event generators become even
more pronounced in the case of antinuclei: The parametrisations rely on
additional approximations like the neglect of two-particle
correlations, while the Monte Carlo description is severely computationally
demanding. This motivated us to extend the interpolation subroutines of
{\tt AAfrag} to include also our predictions
for the production cross sections of antinuclei in $pp$, $p$He, He$p$, HeHe,
$\bar{p}p$, and $\bar{p}$He interactions.

The production of (anti-) nuclei in small interacting systems is arguably best
described using so-called coalescence models. In these models, final-state
nucleons may merge to form a nucleus if they are sufficiently close in phase
space~\cite{Schwarzschild:1963zz,butler_deuterons_1963}.
Currently, the only model that is able to account (semi-classically) for
nucleon momentum correlations in a Monte Carlo framework as well as the nucleon
emission volume is the so-called WiFunC (short for Wigner Functions with
Correlations) model introduced in
Ref.~\cite{Kachelriess:2019taq} and discussed in more detail in
Refs.~\cite{Kachelriess:2020uoh,Kachelriess:2020amp}.
In this model, the probability that a given
(anti-) proton--(anti-) neutron pair coalesce is found by projecting the wave
function describing the final-state nucleons onto the deuteron wave function.
Using for the latter a two-Gaussian wave function, one obtains for the
formation probability $w$ of an (anti-) deuteron
\begin{equation}
  w = 3\Delta \zeta_1 \me^{-d_1^2q^2} + 3(1-\Delta)\zeta_2\me^{-d_2^2q^2},
\end{equation}
where $q$ is the momentum of the nucleons in the pair rest frame and
the parameters $\Delta=0.581$, $d_1=3.979\unit{fm}$ and $d_2=0.890\unit{fm}$
are fixed such that $\phi(r=0)$ and the expectation value $\langle r\rangle$
of the two-Gaussian wave function
have the same values as for the Hulthen wave function~\cite{Zhaba:2017gnm}.
The resulting deuteron yield is then obtained by selecting proton-neutron
pairs obtained in a particle collision in the event generator with
probability $w$.
The suppression factors $\zeta_i$ are given by
\begin{equation}
  \zeta_i^2 = \frac{d_i^2}{d_i^2+4(\sigma m/m_T)^2}
    \frac{d_i^2}{d_i^2+4\sigma^2}\frac{d_i^2}{d_i^2+4\sigma^2},
\end{equation}
and depend on the coalescence parameter\footnote{
  The longitudinal spread is expected to be of the same size
  as the transverse spread. To minimize the number of free parameters
  of the model, we have
  therefore fixed $\sigma\equiv\sigma_\perp=\sigma_\|$. When more experimental
  data and improved event generators become available, the two spreads may have
  to be fitted separately.
} $\sigma$ which determines the size of
the formation region of nucleons; Here, $m$ and $m_T$ are the nucleon mass
and the nucleon transverse mass, respectively. 
The coalescence parameter is expected to be
close to $1\unit{fm}$ and its process dependence can be approximated as
described in Ref.~\cite{Kachelriess:2020uoh}.
Analytical expressions for the coalescence probability $w$ of three
nucleons into (anti-) helium-3
and (anti-) tritium have been derived in Ref.~\cite{Kachelriess:2020uoh},
thereby allowing for a consistent description of the production of 
light nuclei in various interactions
relevant for  astrophysical studies.
It is important to note that this model contains basically no free
parameters: The parameter $\sigma$  can be fixed independently
by femtoscopy experiments. Moreover, the numerical values 
derived for $\sigma$ from  femtoscopy experiments and from fits to 
various production channels of light antinuclei are consistent with
each other and agree with its physical interpetation, $\sigma\simeq 1$\,fm.

In this work, we provide the predictions for the production cross sections
of antideuteron and antihelium-3 in $pp$, $p$He, He$p$, HeHe,
$\bar{p}p$, and $\bar{p}$He interactions, based on the QGSJET-II-04m
Monte Carlo generator and the WiFunC model.
In addition, we comment on the importance of including momentum
correlations when describing the production of astrophysical antinuclei, and on
the interpretation of antideuteron and antihelium experiments at accelerators.
Finally, we argue that the nuclear enhancement in the astrophysically
interesting range is strongly energy dependent and can therefore not be
approximated by a constant factor.

\section{Selected results}

%In Fig.~\ref{fig:alice}, we compare the invariant differential antideuteron
%yield measured by the ALICE collaboration in $pp$ collisions at
%0.9, 2.76 and 7\,TeV~\cite{ALICE:2017xrp} to that obtained using QGSJET-II-04m
% and the
%WiFunC model. This comparison clearly shows that the differential yields are
%well reproduced. One should note, however, that for 2.76 and 7\,TeV,
%QGSJET does not reproduce well the slope of the
%corresponding antiproton spectrum~\cite{Kachelriess:2020uoh}. Therefore, a
%re-weighting\footnote{%
%We emphasise that the re-weighting depends heavily on the collision energy and
%  the kinematical region considered in the experiment. Therefore, it has no
%  predictive power and is not used elsewhere in this work.
%}
%of the antiproton spectrum at these energies has been performed in order to
%obtain a more precise prediction for the coalescence
%factor~\cite{Kachelriess:2020uoh}.

\begin{figure}[htb]
  \centering
  \includegraphics[scale=0.8]{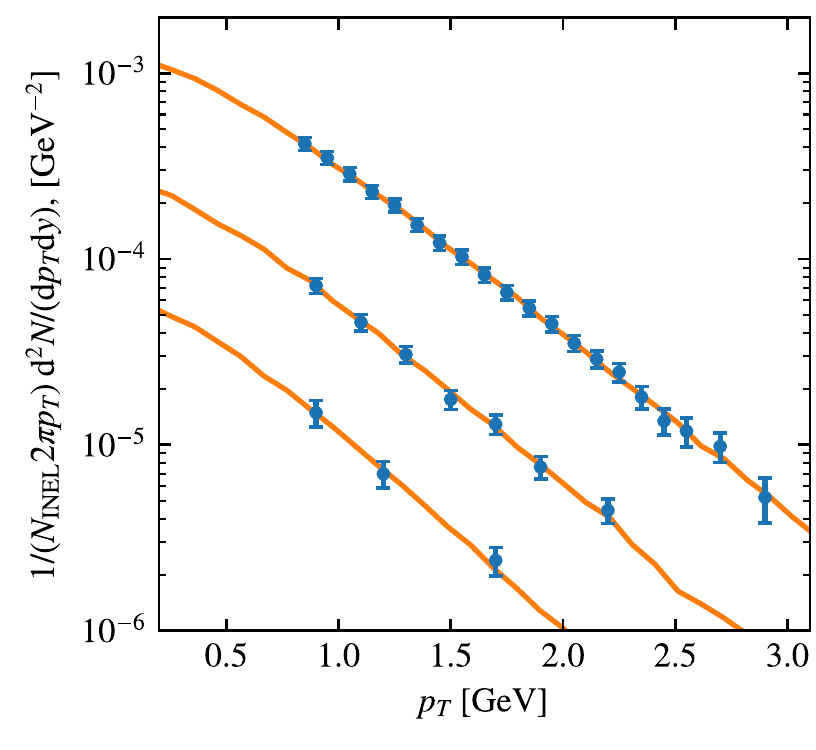}
  \caption{Best fit of the WiFunC model, using QGSJET-II-04m, to the invariant
  differential yield of antideuterons, measured by the ALICE collaboration in
  $pp$ collisions (blue data points) at 0.9, 2.76 and 7 TeV.
  The yields are multiplied by a constant factor to make the figure clearer.
  A re-weighting of the antinucleons was included as in
  Ref.~\cite{Kachelriess:2020uoh}.
  }
  \label{fig:alice}
\end{figure}

In Fig.~\ref{fig:alice}, we compare the invariant differential antideuteron
yield measured by the ALICE collaboration in $pp$ collisions at
0.9, 2.76 and 7\,TeV~\cite{ALICE:2017xrp} to that obtained using QGSJET-II-04m
 and the
WiFunC model. This comparison clearly shows that the differential yields are
well reproduced.

\begin{figure}[htb]
  \centering
  \includegraphics[scale=0.8]{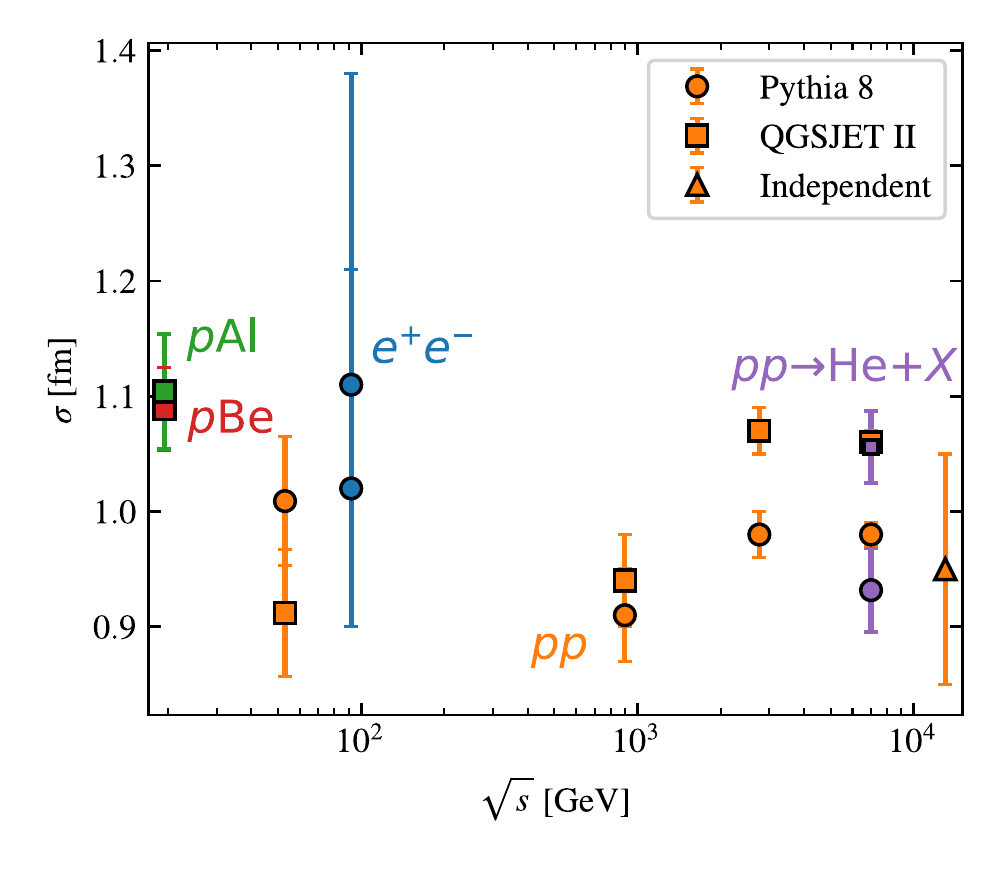}
  \caption{
    A compilation of 
     values of the  coalescence parameter $\sigma$,
    obtained from fits to various
    experimental data on antideuteron and antihelium-3 production in
    $pp$~\cite{ALICE:2017xrp,British-Scandinavian-MIT:1977tan},
    $pN$~\cite{Bozzoli:1979fh},
    and $e^+e^-$~\cite{ALEPH:2006qoi,OPAL:1995uwx}
    collisions using Pythia~8 (circles) and QGSJET-II-04m (squares). The
    triangular data point is obtained from a femtroscopy
    experiment~\cite{Acharya:2020dfb},
    and is thus independent on the event generator. The fitting is explained in
    Ref.~\cite{Tjemsland:2020bzu} and references therein.
  }
  \label{fig:sigma}
\end{figure}

A compilation of fits of $\sigma$ to various accelerator experiments on
antideuteron and antihelium-3 production using QGSJET-II-04m
 and Pythia~8 is shown in
Fig.~\ref{fig:sigma} (see Refs.~\cite{Kachelriess:2020uoh,Kachelriess:2019taq,%
Kachelriess:2020amp,Tjemsland:2020bzu} and references therein for a
discussion of the experimental data and the fitting procedures).
It is clear that the numerical values of $\sigma$ are consistent with being
 constant and equal to $(1.0\pm0.1)$\,fm within the theoretical and
experimental uncertainties. 
It should be emphasised that the triangular data point in Fig.~\ref{fig:sigma}
is obtained from a fit to  the baryon source size and its $m_T$ scaling
measured by the ALICE collaboration in $pp$ collisions~\cite{Acharya:2020dfb}.
Thus this data point represents an independent measurement of the coalesence
parameter $\sigma$, using only data on baryon production. The agreement of this
value with the one found applying the WiFunC model to anti-nuclei production
supports the validity of the basic assumptions underlying this model.
Note also that the free parameter $\sigma$ in the WiFunC model
can be fixed independently of the coalescence mechanism via baryon femtoscopy,
see the last data point in Fig.~2.

%%%%%%%%%%%%%%%%%%%%%%%%%%%%%%%%%%%%%%%%%%%%%%%%%%%%%%%%%%%%%%%%%%%%%%%%%%%%%
\section{Secondary production of antinuclei}

%%%%%%%%%%%%%%%%%%%%%%%%%%%%%%%%%%%%%%%%%%%%%%%%%%%%%%%%%%%%%%%%%%%%%%%%%%%%%
\subsection{Comparison with parametrisation methods}

Traditionally, the production of a nucleus with mass number $A$ has been
parametrised by the proton spectrum as
\begin{equation}
  E_A\dv[3]{N_A}{P_A} = B_A\left(E_N\dv[3]{N_p}{p_p}\right)^{A},
  \label{eq:isotropic}
\end{equation}
where $B_A$ is known as the coalescence factor. In the limit of isotropic
nucleon yields, $B_A$ is a constant that scales with the nucleon emission
volume as $B_A\propto V^{A-1}$ if the coalescence condition is evaluated in
position space, and with the coalescence momentum as $B_A\propto p_0^{3A-3}$
if evaluated in momentum space. In small interacting systems, such as $pp$,
$pN$ and $e^+e^-$ collisions or dark matter annihilations, this
approximation is not valid since the nucleon yield is highly non-isotropic.
Even so, the approximation~\eqref{eq:isotropic} is often used in
astrophysical studies due to its simplicity.

Another reason for deviations from the simple relation~\eqref{eq:isotropic}
is the missing phase-space suppression close to the production threshold.
Since the production channels with the minimal number of particles, compatible
with baryon number conservation, will dominate close to the threshold, one
can approximate the suppression at low collision energies and high secondary 
nucleus energies as a pure phase-space suppression, assuming an isotropic
matrix element (see e.g. Refs.~\cite{Blum:2017qnn,%
Duperray:2002pj,Duperray:2003tv}).
Thus, the approximation~\eqref{eq:isotropic} can be improved if we include
a phase-space suppression factor
\begin{equation}
  R_N(x) \equiv \frac{\Phi_N(x; m_p)}{\Phi_N(x; 0)},
  \label{eq:R}
\end{equation}
where $x=\sqrt{s+Am_p^2-2\sqrt{s}\tilde{E}_A}$ describes the energy
available in the center of mass (CoM) frame, and $\tilde{E}_A$ is the nucleus
energy in the CoM frame.
One can compute the phase-space
integrals,
\begin{equation}
  \Phi_N(x; m_p)=\left[\prod_{i=1}^N\int\frac{\dd[3]{p_i}}{(2\pi)^32E_i}\right]
    (2\pi)^4\delta^{(3)}\left(\sum_{i=1}^{N}\vec{p}_i\right)
    \delta\left(\sum_{i=1}^{N}E_i-x\right),
\end{equation}
with the integration volume being the full allowed phase-space,
numerically using the method described in Ref.~\cite{James:275743}
(see Fig.~15 in Ref.~\cite{Blum:2017qnn} for a plot). This method has a major
perk compared to a Monte Carlo treatment: It is significantly less
computationally demanding. As in the case of the WiFunC model, the method
contains no free parameters since the coalescence factor can be obtained from
femtoscopy experiments~\cite{Blum:2017qnn,Blum:2019suo,Bellini:2020cbj}.
However, the suppression factor is not exact and two-particle correlations are
not taken into account, meaning that one may expect the method to give
inaccurate results. For instance, at low energies near the threshold, the
model is expected to overproduce 
nuclei since it does not take into account anti-correlations.
Furthermore, the coalescence factor $B_A$ is typically determined in a
kinematical regime  not relevant for cosmic ray studies. As such, results
obtained within this approximation have to be interpreted with care.

\begin{figure}
  \centering
  \includegraphics[width=.49\textwidth]{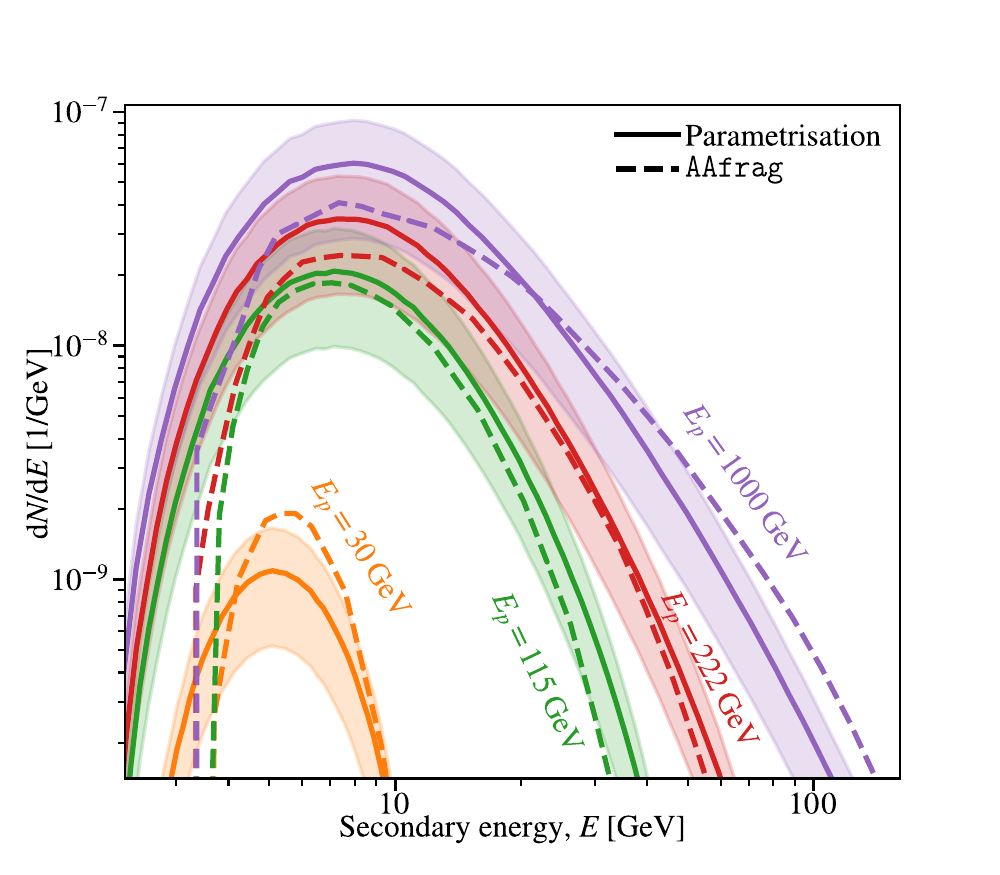}
  \includegraphics[width=.49\textwidth]{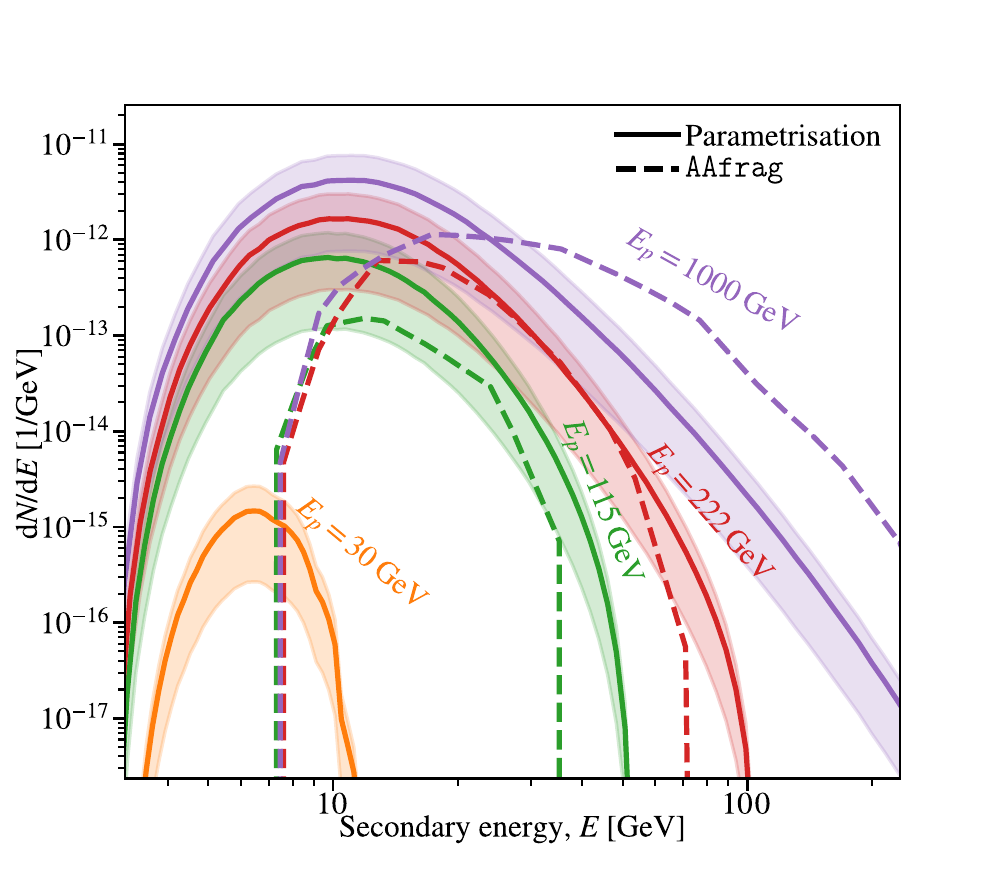}
  \caption{The antideuteron (left) and antihelium-3
    (right) spectra obtained using Eqs.~\eqref{eq:isotropic} and \eqref{eq:R}
    (solid lines) and the WiFunC model (dashed lines). The results are shown
    for primary energies of 21 GeV (blue), 30 GeV (orange), 115 GeV (green),
    222 GeV (red) and 1000 GeV (purple).}
  \label{fig:secondaries}
\end{figure}

In order to verify the importance of using a Monte Carlo description, we
plot in Fig.~\ref{fig:secondaries} the antideuteron (left) and antihelium-3
(right) spectra
${\rm d} N/{\rm d} E={\sigma_{\rm inel}}^{-1}{\rm d}\sigma_{\rm inel}/{\rm d} E$
obtained using Eqs.~\eqref{eq:isotropic} and \eqref{eq:R}
(solid lines) and the WiFunC model (dashed lines) for various primary energies.
The shaded areas correspond to the ranges $B_2=(0.75,2.4)\cdot 10^{-2}$ and
$B_3 = (2,20)\cdot 10^{-3}$ obtained from femtoscopy 
experiments~\cite{Blum:2017qnn}. 
Note that QGSJET-II-04m was used to obtain the antiproton spectrum, and that
using instead
a parametrisation of the antinucleon spectrum would lead to  larger
differences.
While the phase-space suppression factor captures well the overall behaviour
of the suppression for high secondary energies, there are large differences
near the production threshold at low energies. These differences are much larger
for helium-3 than for deuteron. 
A reason for these differences is that the parametrisation fails to
account for the fact that certain processes are kinematically allowed but
forbidden by conservation laws, e.g.\ for baryon number. In particular,
antideuterons may be produced using parametrisations via the (forbidden) process
$pp\to pp\bar{p}\bar{n}$. Furthermore, in high energy collisions and near
the threshold, parametrisations fail to describe the momentum (anti-)
correlations that may enhance or lower the antinuclei
production~\cite{Dal:2012my}.
Although the parametrisation~\eqref{eq:isotropic} describes (within the
experimental and theoretical uncertainties in $B_A$) the overall yield
of antinuclei sufficiently well for order-of-magnitude estimates, an
accurate Monte Carlo description is therefore needed if one aims
to reduce the uncertainties in the theoretical predictions.

%%%%%%%%%%%%%%%%%%%%%%%%%%%%%%%%%%%%%%%%%%%%%%%%%%%%%%%%%%%%%%%%%%%%%%%%%%%%%
\subsection{Nuclear enhancement}

One of the major advantages of the WiFunC model and the use of a Monte Carlo
generator is that one can describe the production of antideuteron and
antihelium-3 in point-like and extended processes without any free parameters.
In particular, one can avoid the use of a  ``nuclear enhancement factor'',
which is otherwise required if the primary and/or the target is a nucleus.
 Previous studies like
those of Refs.~\cite{Lin:2018avl,Ibarra:2013qt} had to assume that the
nuclear enhancement is constant and  coincides with the one for antiproton
production. In fact, these assumptions are invalid, as we shall demonstrate
below.

To discuss the nuclear enhancement of secondary fluxes analytically, it is
convenient to consider the case when the primary cosmic ray spectra can be
approximated by power laws, $I_{A_i}(E_0)\propto E_0^{-\alpha_i}$, with
$\alpha_i$ being the slope and $E_0$ the energy per nucleon. In such a case, the
contribution $q_X^{A_ip}$ to the secondary flux of particles $X$ ($e^{\pm}$,
 $\gamma$, $\nu$, or $\bar p$) from interactions of primary nuclei $A_i$ with
protons from the
interstellar medium is proportional to the weighted moment of the
corresponding production spectrum (see, e.g.~\cite{Kachelriess:2014mga}):
\begin{equation}
  q_X^{A_ip}(E)\propto  Z_X^{A_ip}(E,\alpha_i)\,,
  \label{eq:q_X}
\end{equation}
with
\begin{equation}
  Z_X^{A_ip}(E,\alpha_i)=\int_0^1\!\mathrm{d}z\;z^{\alpha_i-1}\;
  \frac{\mathrm{d}\sigma_{A_ip\rightarrow X}(E/z,z)}{\mathrm{d}z}\,.
  \label{eq:Z_X}
\end{equation}
Then the nuclear enhancement can be quantified by the ratio
$Z_X^{A_ip}(E,\alpha_i)/Z_X^{pp}(E,\alpha_i)$ which would coincide
with the ratio of the respective contributions to the secondary flux of 
interest, $q_X^{A_ip}/q_X^{pp}$, if the primary proton and nuclei flux
would be the same. In the limit of very high energies, one obtains an
$A$ enhancement for that ratio~\cite{Kachelriess:2014mga,Kachelriess:2015wpa},
\begin{equation}
\varepsilon_X^{A_ip}(E)\equiv
Z_X^{A_ip}(E,\alpha_i)/Z_X^{pp}(E,\alpha_i)\rightarrow A_i\,.
  \label{eq:eps_X}
\end{equation}

This simple result follows from two important features of nucleus-proton
(or, more generally, nucleus-nucleus) interactions. First, the forward
(i.e.\ large $z$)
spectrum for any secondary particle $X$ in collisions of a nucleus $A$ with
protons can be approximated by the one for a superposition of 
$\langle n_A^{\rm w}\rangle$ independent $pp$ collisions,
\begin{equation}
\left. \frac{dn_{A_ip\rightarrow X}(E_0,z)}{dz}\right|_{z\rightarrow 1}
\simeq \langle n_A^{\rm w}(E_0)\rangle\,
 \frac{dn_{pp\rightarrow X}(E_0,z)}{dz}\,.
   \label{eq:xf1}
\end{equation}
Here, the average number $\langle n_A^{\rm w}\rangle$ of interacting  
(``wounded'') projectile nucleons satisfies \cite{Bialas1976} 
\begin{equation}
\langle n_A^{\rm w}(E_0)\rangle \simeq \frac{A\,\sigma^{\rm inel}_{pp}(E_0)}
{\sigma^{\rm inel}_{Ap}(E_0)}\,,
  \label{eq:n_w}
\end{equation}
with $\sigma^{\rm inel}_{pp}$ and $\sigma^{\rm inel}_{Ap}$ as the inelastic
cross sections of $pp$ and $Ap$ collisions, respectively. Inserting
Eq.~(\ref{eq:n_w})
into (\ref{eq:xf1}) and substituting the result in (\ref{eq:Z_X}), one arrives
at Eq.~(\ref{eq:eps_X}).

Let us consider now the production of light nuclei. For definiteness,
we will discuss the case of antideuterons. The crucial
difference to the picture described above is that this process proceeds via
the coalescence mechanism and thus involves the double differential spectra of
produced antiprotons and antineutrons,
$\mathrm{d}^2 \sigma_{Ap\rightarrow \bar p+\bar n}/
{\mathrm{d}z_{\bar p} \mathrm{d}z_{\bar n}}$.
In nucleus-proton collisions, the coalescing antiproton and antineutron are 
typically created in rescatterings of different projectile nucleons off the
target proton~\cite{Kachelriess:2020uoh}. Consequently, the forward-production
spectra of antideutrons become proportional to the number of possible  
pair-wise nucleon-proton rescattering processes,
$\propto n_A^{\rm w} (n_A^{\rm w}-1)$. Thus, at sufficiently high energies 
and for large $A$,
the nuclear enhancement for $\bar d$ production should satisfy 
\begin{equation}
\varepsilon_{\bar d}^{Ap}(E)\propto
  \frac{A^2\sigma^{\rm inel}_{pp}(E)}{\sigma^{\rm inel}_{Ap}(E)} \simeq A^{4/3} ,
  \label{eq:eps_dbar}
\end{equation}
with $E$ as the energy per nucleon for $\bar d$. In the last step,
we assumed for illustration a simple  $A^{2/3}$ scaling for the $Ap$ cross
sections.

\begin{figure}[htb]
  \centering
  \includegraphics[height=8cm]{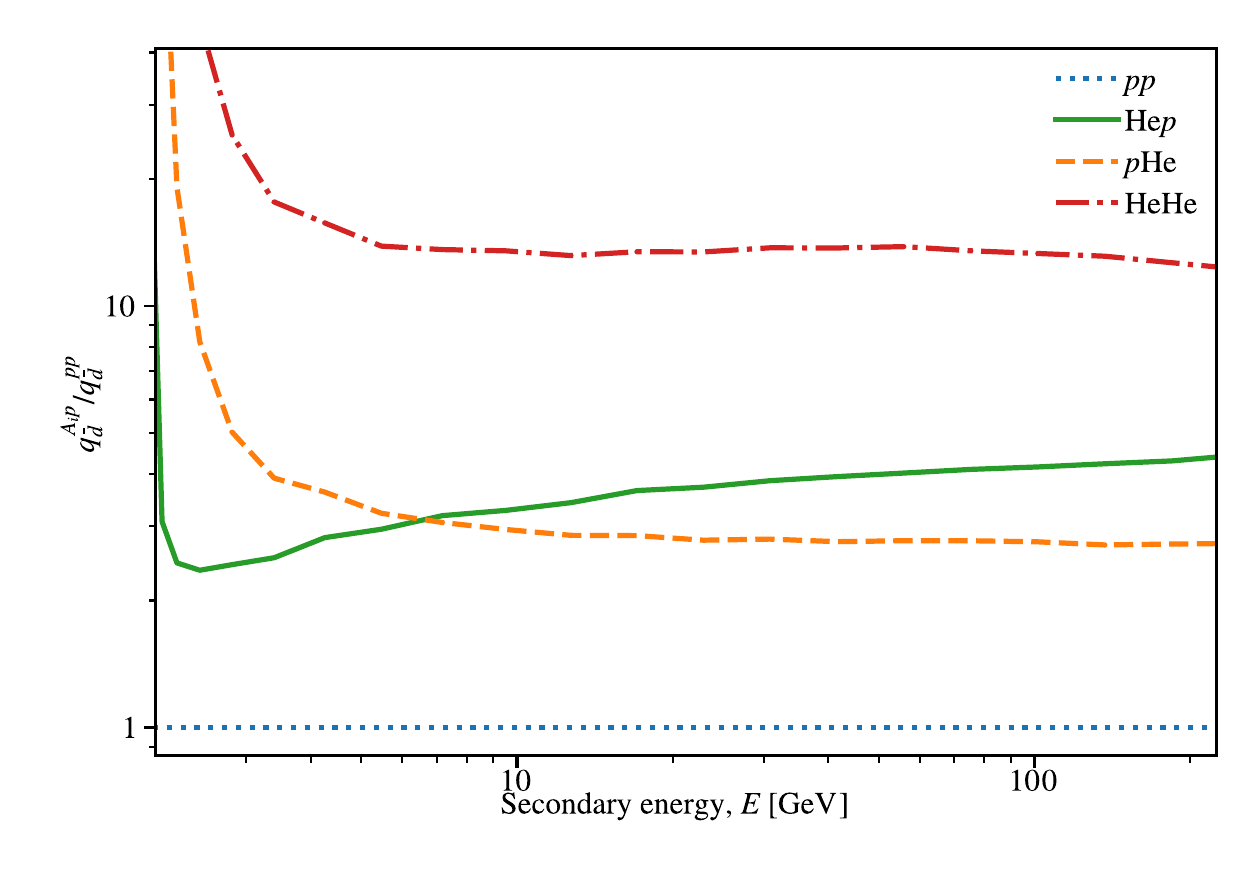}
  \caption{The nuclear enhancement factors
    $\eps_{\bar d}^{AB}(E)=q_X^{A_AB}/q_X^{pp}$ as function of the total
    energy 
for $AB={\rm He}p$, $p{\rm He}$, and ${\rm He}{\rm He}$ collisions.
  \label{fig:enhancement}}
\end{figure}

In Fig.~\ref{fig:enhancement}, we plot the calculated energy dependence of the
enhancement factors $\varepsilon_{\bar d}^{AB}(E)=q_X^{AB}/q_X^{pp}$,
for $AB={\rm He}p$, $p{\rm He}$, and ${\rm He}{\rm He}$ and assuming
the same primary proton and helium flux. At the highest energies,
$\varepsilon_{\bar d}^{{\rm He}p}$ is larger than four and increasing
towards the asymptotic limit, Eq.~(\ref{eq:eps_dbar}).
In contrast, the behaviour
of $\varepsilon_{\bar d}^{AB}(E)$
changes drastically in the low-energy limit. Such a trend has been previously
observed and explained in Ref.~\cite{Kachelriess:2015wpa}, for the case of
$\bar p$ production: Because of the relatively high proton mass, at low energies
the integral in Eq.~(\ref{eq:Z_X}) is no longer dominated by forward (large
$z$)  $\bar p$ production. Instead, important contributions come from the
central (in the center of mass frame) and backward production, such that the
reasoning which lead to Eq.~(\ref{eq:eps_X}) becomes inapplicable.
The same considerations fully apply to the case of antideutron production.
Moreover, regarding antideutron production in proton-helium collisions,
the coalescing $\bar p$ and $\bar n$ are predominantly produced  in inelastic 
rescatterings on different target nucleons. As a result, the  energy threshold
for $\bar d$ production is lower than in $pp$ interactions, and gives rise to
a large nuclear enhancement close to the production threshold.

%%%%%%%%%%%%%%%%%%%%%%%%%%%%%%%%%%%%%%%%%%%%%%%%%%%%%%%%%%%%%%%%%%%%%%%%%%%%%
\subsection{The use of event generators and the interpretation of accelerator
data}

Currently, the best experimental data on antinuclei production, e.g.\ from
the ALICE experiment, are obtained at  energies and in kinematical regions
that are not relevant for astrophysical studies.
Fitting phenomenological coalescence models to such data
leads to a biased model with a reduced predictive power.
Therefore it is important to always assess the applicability of an event
generator, e.g., by comparing with antiproton data obtained under the same
conditions, when comparing the coalescence model to experimental data.
For example, QGSJET does not reproduce well the slope of the antiproton spectrum
at 2.76 and 7\,TeV measured by ALICE~\cite{Kachelriess:2020uoh}. Therefore, a
re-weighting\footnote{%
We emphasise that the re-weighting depends heavily on the collision energy and
  the kinematical region considered in the experiment. Therefore, it has no
  predictive power and is not used elsewhere in this work.
}
of the antiproton spectrum at these energies has been performed in order to
obtain a more precise prediction for the coalescence
factor~\cite{Kachelriess:2020uoh} in Fig.~\ref{fig:alice}.
If QGSJET-II-04m is blindly applied
to the 7\,TeV ALICE data, the obtained value for the coalescence parameter
$\sigma$ is 1.4\,fm, while adjusting the results to the antiproton data yields
1.1\,fm~\cite{Kachelriess:2020uoh,Tjemsland:2020bzu}.

%%%%%%%%%%%%%%%%%%%%%%%%%%%%%%%%%%%%%%%%%%%%%%%%%%%%%%%%%%%%%%%%%%%%%%%%%%%%%
\section{Program structure and example output}

For convenience of both the young generation and ancient users,
the program {\tt AAfrag 2} exists as both Python 3 and Fortran 90 versions.

%%%%%%%%%%%%%%%%%%%%%%%%%%%%%%%%%%%%%%%%%%%%%%%%%%%%%%%%%%%%%%%%%%%%%%%%%%%%%
\subsection{Purpose and method}

{\tt Aafrag 2} is a tool that interpolates results
relevant for secondary interactions in cosmic ray studies from
the Monte Carlo simulation QGSJET-II-04m. The calculation of the production
cross section of photons, neutrinos, electrons, positrons,
(anti-) protons and (anti-) neutrons in $pp$, $p$He, He$p$,
HeHe, C$p$, Al$p$, Fe$p$ interactions,
as well as production cross section of antideuteron and
antihelium-3 
in $pp$, $p$He, He$p$, HeHe, $\bar{p}p$ and $\bar{p}$He interactions,
are included.
The tool allows the users to benefit from the advantages of a Monte Carlo
simulation, with minimal computational effort.
The calculations of photons, neutrinos, electrons, positrons,
(anti-) protons and (anti-) neutrons  were discussed in
Ref.~\cite{Kachelriess:2019ifk}, while the
cases of antideuteron and antihelium were treated in this work.

The results from the Monte Carlo simulations were stored in
tables, and the main purpose of {\tt AAfrag 2} is to provide the user with
convenient interpolation routines. The interpolation is performed using
bilinear interpolation, with a fill value 0 outside the data range.

The Fortran 90 program includes its own interpolation routines and is thus
self-consistent, while the Python 3 program depends
on the {\tt numpy}, {\tt scipy} and {\tt matplotlib} libraries.

\subsection{Program structure}

The Fortran 90 program consists of two Fortran files,
{\tt AAfrag2.f90} and {\tt user.f90},
in addition to the numerical tables in the {\tt Tables} folder. The file 
{\tt AAfrag2.f90} contains the module {\tt spectra} which is used to store
physical parameters and the loaded tables, the main program, subroutines used to
initialise and load the tables, and the interpolation functions. Meanwhile,
the file {\tt user.f90} contains an example calculation. For the normal user,
only changes in {\tt user.f90} are necessary.
The main program calls
{\tt init} which loads all the tables and stores the data in the variables
defined in the module {\tt spectra}. Next, the subroutine {\tt user\_main} in
{\tt user.f90} is called. Users must adapt this subroutine to their
specific needs.

The Python 3 program is contained in
{\tt AAfrag2.py}, in addition to the numerical tables in the
{\tt Tables} folder. The Python functions
have the same names and input parameters as the Fortran subroutines and
functions. The script contains an example calculation that is executed when the
script is run as a standalone. The user can either change the example portion of
the script, or import the file as a module.

\subsection{Functions}
\label{sec:functions}

The program includes five functions that are intended to be used by the user
({\tt spec\_gam}, {\tt spec\_nu}, {\tt spec\_elpos}, {\tt spec\_pap},
{\tt spec\_nan}, {\tt spec\_ad}, {\tt spec\_ah})
which interpolate the production spectra of
secondaries $\{\gamma,$ $\nu_i,$ $e^-,$ $e^+,$ $p,$ $\bar{p},$ $n,$ $\bar{n},$
$\bar{d}$, $^3\bar{\rm He}$+$^3\bar{\rm H}\}$ in various cosmic ray
interactions. The functions have the same input parameters:
{\tt (E\_p, E\_s, q, k)}. Here, {\tt E\_p} is the total energy of the
primary nucleus in GeV, {\tt E\_s} is the kinetic energy of the produced
secondary in GeV, {\tt q} denotes the particle species as detailed in
Tab.~\ref{tab:q}, and {\tt k} denotes the interaction type as detailed in
Tab.~\ref{tab:k}. The output is the production spectrum
$E_s^2\dv*{\sigma^k(E_p,E_s)}{E_s}$ in GeV\,mb.
An example of their uses is given in the example programs.

\begin{table}
  \centering
  \begin{tabular}{lcccc}
    \hline
    Function          & $q=1$   & $q=2$         & $q=3$     & $q=4$           \\
    \hline
    {\tt spec\_nu}    & $\nu_e$ & $\bar{\nu}_e$ & $\nu_\mu$ & $\bar{\nu}_\mu$ \\
    {\tt spec\_elpos} & $e^-$   & $e^+$         &    --     &    --           \\
    {\tt spec\_pap}   & $p$     & $\bar{p}$     &    --     &    --           \\
    {\tt spec\_nan}   & $n$     & $\bar{n}$     &    --     &    --           \\
    {\tt spec\_gam}   & $\gamma$ &     --       &    --     &    --           \\
    {\tt spec\_ad}    & $\bar{d}$ &    --       &    --     &    --           \\
    {\tt spec\_ah}    & $^3\bar{\rm He}+{^3}\bar{\rm H}$ & -- & -- & --       \\
    \hline
  \end{tabular}
  \caption{The particle type determined by the parameter $q$.}
  \label{tab:q}
\end{table}

\begin{table}
  \centering
  \begin{tabular}{lccccccccc}
    \hline
                  & $k=1$ & $k=2$ & $k=3$ & $k=4$ & $k=5$ & $k=6$ & $k=7$
                  & $k=8$ & $k=9$           \\
    \hline
    Beam--Target&$p$--$p$&$p$--He&He--$p$&He--He&C--$p$&Al--$p$&Fe--$p$&
                $\bar{p}$--$p$&$\bar{p}$--He\\
    Mass number & 1--1&1--4&4--1&4--4&12--1&26--1&56--1&1--1&1--4 \\
    \hline
  \end{tabular}
  \caption{Reaction type determined by the parameter $k$. Reactions 5--7 are
  not implemented for antideuteron ({\tt spec\_ad}) and antihelium-3
  ({\tt spec\_ah}), while reactions 8--9 are exclusively implemented for them.}
  \label{tab:k}
\end{table}

%%%%%%%%%%%%%%%%%%%%%%%%%%%%%%%%%%%%%%%%%%%%%%%%%%%%%%%%%%%%%%%%%%%%%%%%%%%%%
\subsection{Example output}

\begin{figure}[htb]
  \centering
  \includegraphics[height=8cm]{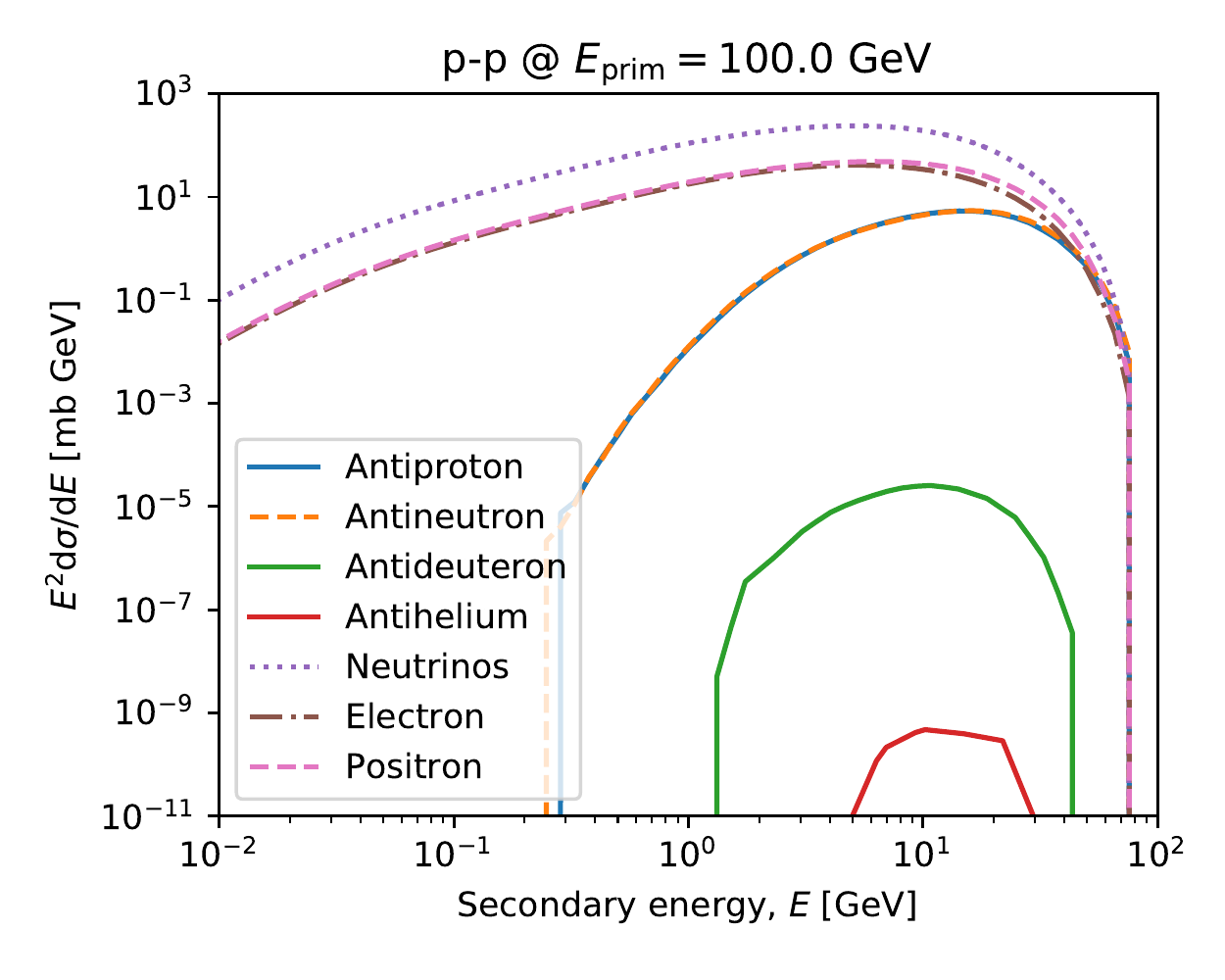}
  \caption{Production spectrum of antiprotons, antineutrons, antideuterons,
  neutrinos, positrons and electrons in $pp$ collisions with primary energy
  of 100 GeV.
  The figure is obtained by running the example programs included in
  {\tt AAfrag2}.}
  \label{fig:example}
\end{figure}

The output of the example calculation are the files {\tt spec\_gam},
{\tt spec\_nu}, {\tt spec\_elpos}, {\tt spec\_aprot}, {\tt spec\_aneut},   
{\tt spec\_adeut} and {\tt spec\_ahel}. They contain respectively the production
spectra of photons, neutrinos, electrons and positrons, antiprotons,
antineutrons, antideuterons, and antihelium-3. The result is plotted in
Fig.~\ref{fig:example} for $pp$ collisions at 100 GeV. 

%%%%%%%%%%%%%%%%%%%%%%%%%%%%%%%%%%%%%%%%%%%%%%%%%%%%%%%%%%%%%%%%%%%%%%%%%%%%%
\section{Summary}

Astrophysical antideuteron and antihelium-3 are ideal probes for new, exotic
physics due to the suppressed background at low energies. Therefore, we made our
predictions for the production cross sections of 
antideuteron and antihelium-3 in $pp$, $p$He, He$p$, HeHe, $\bar pp$ and
$\bar p$He collisions at energies relevant for secondary production in the
Galaxy publicly available through the interpolation
subroutines {\tt AAfrag}.
The predictions are based on QGSJET-II-04m and the state of the art coalescence
model evaluated on an event-by-event basis. 
Furthermore, we commented on the use of Monte Carlo generators to predict 
antinuclei fluxes, the use of a nuclear enhancement factor to predict secondary
cosmic ray fluxes, and the effect of two-particle
momentum correlations provided by a Monte Carlo.

%%%%%%%%%%%%%%%%%%%%%%%%%%%%%%%%%%%%%%%%%%%%%%%%%%%%%%%%%%%%%%%%%%%%%%%%%%%%%
\section{Acknowledgements}

This research was supported by the Munich Institute for Astro- and Particle
Physics (MIAPP) which is funded by the Deutsche Forschungsgemeinschaft (DFG,
German Research Foundation) under Germany's Excellence Strategy -- EXC-2094 --
390783311. S.O. acknowledges support from the Deutsche
Forschungsgemeinschaft (project number 465275045).

%\bibliographystyle{elsarticle-num}
%\bibliography{references}

\begin{thebibliography}{10}
\expandafter\ifx\csname url\endcsname\relax
  \def\url#1{\texttt{#1}}\fi
\expandafter\ifx\csname urlprefix\endcsname\relax\def\urlprefix{URL }\fi
\expandafter\ifx\csname href\endcsname\relax
  \def\href#1#2{#2} \def\path#1{#1}\fi

\bibitem{Neronov:2011wi}
A.~Neronov, D.~V. Semikoz, A.~M. Taylor, {Low-energy break in the spectrum of
  Galactic cosmic rays}, Phys. Rev. Lett. 108 (2012) 051105.
\newblock \href {http://arxiv.org/abs/1112.5541} {\path{arXiv:1112.5541}},
  \href {https://doi.org/10.1103/PhysRevLett.108.051105}
  {\path{doi:10.1103/PhysRevLett.108.051105}}.

\bibitem{Fermi-LAT:2012edv}
M.~Ackermann, et~al., {Fermi-LAT Observations of the Diffuse Gamma-Ray
  Emission: Implications for Cosmic Rays and the Interstellar Medium},
  Astrophys. J. 750 (2012) 3.
\newblock \href {http://arxiv.org/abs/1202.4039} {\path{arXiv:1202.4039}},
  \href {https://doi.org/10.1088/0004-637X/750/1/3}
  {\path{doi:10.1088/0004-637X/750/1/3}}.

\bibitem{Kachelriess:2012fz}
M.~Kachelrie{\ss}, S.~Ostapchenko, {Deriving the cosmic ray spectrum from
  gamma-ray observations}, Phys. Rev. D 86 (2012) 043004.
\newblock \href {http://arxiv.org/abs/1206.4705} {\path{arXiv:1206.4705}},
  \href {https://doi.org/10.1103/PhysRevD.86.043004}
  {\path{doi:10.1103/PhysRevD.86.043004}}.

\bibitem{Donato:1999gy}
F.~Donato, N.~Fornengo, P.~Salati, {Anti-deuterons as a signature of
  supersymmetric dark matter}, Phys. Rev. D62 (2000) 043003.
\newblock \href {http://arxiv.org/abs/hep-ph/9904481}
  {\path{arXiv:hep-ph/9904481}}, \href
  {https://doi.org/10.1103/PhysRevD.62.043003}
  {\path{doi:10.1103/PhysRevD.62.043003}}.

\bibitem{vonDoetinchem:2020vbj}
P.~von Doetinchem, et~al., {Cosmic-ray Antinuclei as Messengers of New Physics:
  Status and Outlook for the New Decade}, JCAP 08 (2020) 035.
\newblock \href {http://arxiv.org/abs/2002.04163} {\path{arXiv:2002.04163}},
  \href {https://doi.org/10.1088/1475-7516/2020/08/035}
  {\path{doi:10.1088/1475-7516/2020/08/035}}.

\bibitem{Kachelriess:2014oma}
M.~Kachelrie\ss{}, S.~Ostapchenko, {Neutrino yield from Galactic cosmic rays},
  Phys. Rev. D 90~(8) (2014) 083002.
\newblock \href {http://arxiv.org/abs/1405.3797} {\path{arXiv:1405.3797}},
  \href {https://doi.org/10.1103/PhysRevD.90.083002}
  {\path{doi:10.1103/PhysRevD.90.083002}}.

\bibitem{Kachelriess:2014mga}
M.~Kachelrie{\ss}, I.~V. Moskalenko, S.~S. Ostapchenko, {Nuclear enhancement of
  the photon yield in cosmic ray interactions}, Astrophys. J. 789 (2014) 136.
\newblock \href {http://arxiv.org/abs/1406.0035} {\path{arXiv:1406.0035}},
  \href {https://doi.org/10.1088/0004-637X/789/2/136}
  {\path{doi:10.1088/0004-637X/789/2/136}}.

\bibitem{Kachelriess:2020uoh}
M.~Kachelrie{\ss}, S.~Ostapchenko, J.~Tjemsland, {Revisiting cosmic ray
  antinuclei fluxes with a new coalescence model}, JCAP 08 (2020) 048.
\newblock \href {http://arxiv.org/abs/2002.10481} {\path{arXiv:2002.10481}},
  \href {https://doi.org/10.1088/1475-7516/2020/08/048}
  {\path{doi:10.1088/1475-7516/2020/08/048}}.

\bibitem{Kachelriess:2019ifk}
M.~Kachelrie\ss{}, I.~V. Moskalenko, S.~Ostapchenko, {AAfrag: Interpolation
  routines for Monte Carlo results on secondary production in proton-proton,
  proton-nucleus and nucleus-nucleus interactions}, Comput. Phys. Commun. 245
  (2019) 106846.
\newblock \href {http://arxiv.org/abs/1904.05129} {\path{arXiv:1904.05129}},
  \href {https://doi.org/10.1016/j.cpc.2019.08.001}
  {\path{doi:10.1016/j.cpc.2019.08.001}}.

\bibitem{Ostapchenko:2010vb}
S.~Ostapchenko, {Monte Carlo treatment of hadronic interactions in enhanced
  Pomeron scheme: I. QGSJET-II model}, Phys. Rev. D83 (2011) 014018.
\newblock \href {http://arxiv.org/abs/1010.1869} {\path{arXiv:1010.1869}},
  \href {https://doi.org/10.1103/PhysRevD.83.014018}
  {\path{doi:10.1103/PhysRevD.83.014018}}.

\bibitem{Ostapchenko:2013pia}
S.~Ostapchenko, {QGSJET-II: physics, recent improvements, and results for air
  showers}, EPJ Web Conf. 52 (2013) 02001.
\newblock \href {https://doi.org/10.1051/epjconf/20125202001}
  {\path{doi:10.1051/epjconf/20125202001}}.

\bibitem{Kachelriess:2015wpa}
M.~Kachelrie{\ss}, I.~V. Moskalenko, S.~S. Ostapchenko, {New calculation of
  antiproton production by cosmic ray protons and nuclei}, Astrophys. J.
  803~(2) (2015) 54.
\newblock \href {http://arxiv.org/abs/1502.04158} {\path{arXiv:1502.04158}},
  \href {https://doi.org/10.1088/0004-637X/803/2/54}
  {\path{doi:10.1088/0004-637X/803/2/54}}.

\bibitem{Koldobskiy:2021nld}
S.~Koldobskiy, M.~Kachelrie\ss{}, A.~Lskavyan, A.~Neronov, S.~Ostapchenko,
  D.~V. Semikoz, {Energy spectra of secondaries in proton-proton interactions},
  Phys. Rev. D 104~(12) (2021) 123027.
\newblock \href {http://arxiv.org/abs/2110.00496} {\path{arXiv:2110.00496}},
  \href {https://doi.org/10.1103/PhysRevD.104.123027}
  {\path{doi:10.1103/PhysRevD.104.123027}}.

\bibitem{Schwarzschild:1963zz}
A.~Schwarzschild, C.~Zupancic, {Production of Tritons, Deuterons, Nucleons, and
  Mesons by 30-GeV Protons on A-1, Be, and Fe Targets}, Phys. Rev. 129 (1963)
  854--862.
\newblock \href {https://doi.org/10.1103/PhysRev.129.854}
  {\path{doi:10.1103/PhysRev.129.854}}.

\bibitem{butler_deuterons_1963}
S.~T. Butler, C.~A. Pearson, Deuterons from high-energy proton bombardment of
  matter, Phys. Rev. 129~(2) (1963) 836--842.
\newblock \href {https://doi.org/10.1103/PhysRev.129.836}
  {\path{doi:10.1103/PhysRev.129.836}}.

\bibitem{Kachelriess:2019taq}
M.~Kachelrie{\ss}, S.~Ostapchenko, J.~Tjemsland, {Alternative coalescence model
  for deuteron, tritium, helium-3 and their antinuclei}, Eur. Phys. J. A56~(1)
  (2020) 4.
\newblock \href {http://arxiv.org/abs/1905.01192} {\path{arXiv:1905.01192}},
  \href {https://doi.org/10.1140/epja/s10050-019-00007-9}
  {\path{doi:10.1140/epja/s10050-019-00007-9}}.

\bibitem{Kachelriess:2020amp}
M.~Kachelrie{\ss}, S.~Ostapchenko, J.~Tjemsland, {On nuclear coalescence in
  small interacting systems}, Eur. Phys. J. A 57~(5) (2021) 167.
\newblock \href {http://arxiv.org/abs/2012.04352} {\path{arXiv:2012.04352}},
  \href {https://doi.org/10.1140/epja/s10050-021-00469-w}
  {\path{doi:10.1140/epja/s10050-021-00469-w}}.

\bibitem{Zhaba:2017gnm}
V.~I. Zhaba, {Deuteron: analytical forms of wave function and density
  distribution}, APS Physics 42 (2017) 191--195.
\newblock \href {http://arxiv.org/abs/1802.02778} {\path{arXiv:1802.02778}},
  \href {https://doi.org/10.24144/2415-8038.2017.42.191-195}
  {\path{doi:10.24144/2415-8038.2017.42.191-195}}.

\bibitem{Tjemsland:2020bzu}
J.~Tjemsland, {Formation of light (anti)nuclei}, PoS TOOLS2020 (2021) 006.
\newblock \href {http://arxiv.org/abs/2012.12252} {\path{arXiv:2012.12252}},
  \href {https://doi.org/10.22323/1.392.0006} {\path{doi:10.22323/1.392.0006}}.

\bibitem{Acharya:2020dfb}
S.~Acharya, et~al., {Search for a common baryon source in high-multiplicity pp
  collisions at the LHC}, Phys. Lett. B 811 (2020) 135849.
\newblock \href {http://arxiv.org/abs/2004.08018} {\path{arXiv:2004.08018}},
  \href {https://doi.org/10.1016/j.physletb.2020.135849}
  {\path{doi:10.1016/j.physletb.2020.135849}}.

\bibitem{ALICE:2017xrp}
S.~Acharya, et~al., {Production of deuterons, tritons, $^{3}$He nuclei and
  their antinuclei in pp collisions at $\mathbf{\sqrt{{\textit s}}}$ = 0.9,
  2.76 and 7 TeV}, Phys. Rev. C 97~(2) (2018) 024615.
\newblock \href {http://arxiv.org/abs/1709.08522} {\path{arXiv:1709.08522}},
  \href {https://doi.org/10.1103/PhysRevC.97.024615}
  {\path{doi:10.1103/PhysRevC.97.024615}}.

\bibitem{British-Scandinavian-MIT:1977tan}
S.~Henning, et~al., {Production of Deuterons and anti-Deuterons in Proton
  Proton Collisions at the CERN ISR}, Lett. Nuovo Cim. 21 (1978) 189.
\newblock \href {https://doi.org/10.1007/BF02822248}
  {\path{doi:10.1007/BF02822248}}.

\bibitem{Bozzoli:1979fh}
W.~Bozzoli, A.~Bussiere, G.~Giacomelli, E.~Lesquoy, R.~Meunier, L.~Moscoso,
  A.~Muller, D.~E. Plane, F.~Rimondi, S.~Zylberajch, {Search for Longlived
  Particles in 200-{GeV}/$c$ Proton - Nucleon Collisions}, Nucl. Phys. B 159
  (1979) 363--382.
\newblock \href {https://doi.org/10.1016/0550-3213(79)90340-7}
  {\path{doi:10.1016/0550-3213(79)90340-7}}.

\bibitem{ALEPH:2006qoi}
S.~Schael, et~al., {Deuteron and anti-deuteron production in e+ e- collisions
  at the Z resonance}, Phys. Lett. B 639 (2006) 192--201.
\newblock \href {http://arxiv.org/abs/hep-ex/0604023}
  {\path{arXiv:hep-ex/0604023}}, \href
  {https://doi.org/10.1016/j.physletb.2006.06.043}
  {\path{doi:10.1016/j.physletb.2006.06.043}}.

\bibitem{OPAL:1995uwx}
R.~Akers, et~al., {Search for heavy charged particles and for particles with
  anomalous charge in $e^{+} e^{-}$ collisions at LEP}, Z. Phys. C 67 (1995)
  203--212.
\newblock \href {https://doi.org/10.1007/BF01571281}
  {\path{doi:10.1007/BF01571281}}.

\bibitem{Blum:2017qnn}
K.~Blum, K.~C.~Y. Ng, R.~Sato, M.~Takimoto, {Cosmic rays, antihelium, and an
  old navy spotlight}, Phys. Rev. D96~(10) (2017) 103021.
\newblock \href {http://arxiv.org/abs/1704.05431} {\path{arXiv:1704.05431}},
  \href {https://doi.org/10.1103/PhysRevD.96.103021}
  {\path{doi:10.1103/PhysRevD.96.103021}}.

\bibitem{Duperray:2002pj}
R.~P. Duperray, K.~V. Protasov, A.~Y. Voronin, {Anti-deuteron production in
  proton proton and proton nucleus collisions}, Eur. Phys. J. A 16 (2003)
  27--34.
\newblock \href {http://arxiv.org/abs/nucl-th/0209078}
  {\path{arXiv:nucl-th/0209078}}, \href
  {https://doi.org/10.1140/epja/i2002-10074-0}
  {\path{doi:10.1140/epja/i2002-10074-0}}.

\bibitem{Duperray:2003tv}
R.~P. Duperray, K.~V. Protasov, L.~Derome, M.~Buenerd, {A Model for A = 3
  anti-nuclei production in proton nucleus collisions}, Eur. Phys. J. A 18
  (2003) 597--604.
\newblock \href {http://arxiv.org/abs/nucl-th/0301103}
  {\path{arXiv:nucl-th/0301103}}, \href
  {https://doi.org/10.1140/epja/i2003-10099-9}
  {\path{doi:10.1140/epja/i2003-10099-9}}.

\bibitem{James:275743}
F.~E. James, {Monte Carlo phase space}, CERN, Geneva, 1968, p. 41 p, [numerical
  implementation given in the GENBOD subroutine (W515) in the CERNLIB Fortran
  libraries].
\newblock \href {https://doi.org/10.5170/CERN-1968-015}
  {\path{doi:10.5170/CERN-1968-015}}.

\bibitem{Blum:2019suo}
K.~Blum, M.~Takimoto, {Nuclear coalescence from correlation functions}, Phys.
  Rev. C 99~(4) (2019) 044913.
\newblock \href {http://arxiv.org/abs/1901.07088} {\path{arXiv:1901.07088}},
  \href {https://doi.org/10.1103/PhysRevC.99.044913}
  {\path{doi:10.1103/PhysRevC.99.044913}}.

\bibitem{Bellini:2020cbj}
F.~Bellini, K.~Blum, A.~P. Kalweit, M.~Puccio, {Examination of coalescence as
  the origin of nuclei in hadronic collisions}, Phys. Rev. C 103~(1) (2021)
  014907.
\newblock \href {http://arxiv.org/abs/2007.01750} {\path{arXiv:2007.01750}},
  \href {https://doi.org/10.1103/PhysRevC.103.014907}
  {\path{doi:10.1103/PhysRevC.103.014907}}.

\bibitem{Dal:2012my}
L.~A. Dal, M.~Kachelrie{\ss}, {Antideuterons from dark matter annihilations and
  hadronization model dependence}, Phys. Rev. D86 (2012) 103536.
\newblock \href {http://arxiv.org/abs/1207.4560} {\path{arXiv:1207.4560}},
  \href {https://doi.org/10.1103/PhysRevD.86.103536}
  {\path{doi:10.1103/PhysRevD.86.103536}}.

\bibitem{Lin:2018avl}
S.-J. Lin, X.-J. Bi, P.-F. Yin, {Expectations of the Cosmic Antideuteron Flux}
  (2018).
\newblock \href {http://arxiv.org/abs/1801.00997} {\path{arXiv:1801.00997}}.

\bibitem{Ibarra:2013qt}
A.~Ibarra, S.~Wild, {Determination of the Cosmic Antideuteron Flux in a Monte
  Carlo approach}, Phys. Rev. D 88 (2013) 023014.
\newblock \href {http://arxiv.org/abs/1301.3820} {\path{arXiv:1301.3820}},
  \href {https://doi.org/10.1103/PhysRevD.88.023014}
  {\path{doi:10.1103/PhysRevD.88.023014}}.

\bibitem{Bialas1976}
A.~{Bia{\l}as}, M.~{Bleszynski}, W.~{Czyz}, {}, \npb 111 (1976) 461.

\end{thebibliography}

\end{document}